\documentclass[a4paper,aps,pre,showpacs,twocolumn,superscriptaddress]{revtex4}
\usepackage{amssymb}

\usepackage{graphicx}

\begin{document}

\title{Complex structures in generalized small worlds }
\author{Marcelo Kuperman}
\email{kuperman@cab.cnea.gov.ar}
\affiliation{Centro At{\'o}mico Bariloche and Instituto Balseiro, 8400 S. C.
de Bariloche, Argentina}

\author{Guillermo Abramson}
\email{abramson@cab.cnea.gov.ar}
\affiliation{Centro At{\'o}mico Bariloche and Instituto Balseiro, 8400 S. C.
de Bariloche, Argentina}

\affiliation{Consejo Nacional de Investigaciones Cient{\'\i}ficas y T{\'e}cnicas,
Argentina}

\date{\today}

\begin{abstract}
We propose a generalization of small world networks, in which the
reconnection of links is governed by a function that depends on
the distance between the elements to be linked. An adequate
choice of this function lets us control the clusterization of the
system. Control of the clusterization, in turn, allows the
generation of a wide variety of topologies.
\end{abstract}


\maketitle


\section{Introduction}

The concept of small world was introduced by Milgram
\cite{milgram} in order to describe the topological properties of
social communities and relationships. Recently, a model of these
has been introduced through what was called small world (SW)
networks \cite{watts98}. In the original model of SW networks a
single parameter $p$, running from 0 to 1, characterizes the
degree of disorder of the network, respectively ranging from a
regular lattice to a completely random graph \cite{watts2}. The
construction of these networks starts from a regular,
one-dimensional, periodic lattice of $N$ elements and coordination
number $2K$. Then each of the sites is visited, rewiring $K$ of
its links with probability $p$. Values of $p$ within the interval
$[0,1]$ produce a continuous spectrum of small world networks.
Note that $p$ is the fraction of modified regular links. To
characterize the topological properties of the small world
networks two magnitudes are calculated \cite{watts98}. The first
one, $L(p)$, measures the mean distance between any pair of
elements in the network, that is, the shortest path between two
vertices, averaged over all pairs of vertices. Thus, an ordered
lattice has $L(0)\sim N/K$, while, for a random network,
$L(1)\sim \ln (N)/\ln (K)$. The second one, $C(p)$, measures the
mean clustering of an element's neighborhood . $C(p)$ is defined
in the following way: let's consider the element $i$, having
$k_i$ neighbors connected to it. We call $c_i(p)$ the number of
neighbors of element $i$ that are neighbors among themselves,
normalized to the value that this would have if all of them were
connected to one another; namely $k_{i}(k_{i}-1)/2$. Now $C(p)$
is the average, over the system, of the local clusterization $
c_{i}(p)$. Ordered lattices are highly clustered, with $C(0)\sim
3/4$, and random lattices are characterized by $C(1)\sim K/N$.
Between these extremes small-worlds are characterized by a short
length between elements, like random networks, and high
clusterization, like ordered ones.

Other procedures for developing social networks have been proposed
\cite {ama,albe}. There, an evolving network is considered. The
network grows by the addition of nodes and links. The new links
are added according to the \textit{sociability} or
\textit{popularity} of the individual it connects to. That is, on
the evolutionary growth of the network, those nodes with a higher
number of links are granted with new links with the highest
probability. This process gives place to topologically different
networks characterized by the site connectivity distribution
($\delta$). In \cite{ama} three different topologies are shown:
scale free networks where $\delta$ decays as a power law, broad
scale networks, with $\delta$ behaving as a power law and a
cutoff and single scale networks with $\delta$ having a fast
decaying tail. On the other side in \cite{albe}, in addition to
the scale free networks, the case with $\delta$ following an
exponential is presented.

In the present work, we propose that the reconnection of a link
be governed by a distribution $\Theta$, depending on the distance
between nodes instead of its `popularity.' By changing the
probability distribution $\Theta$ we can obtain topologically
different SW networks, favoring the preservation of high
clusterization. This is a statistical or macroscopical magnitude.
Observing the microscopical properties of the networks a deeper
knowledge of what happens is acquired.

\section{Construction procedure}

As mentioned in the introduction, we perform a generalization of
the original SW networks construction \cite{watts2}. As such, the
SW we study are random networks built upon a topological ring with
$N$ vertices and coordination number $2K$. Each link connecting a
vertex to a neighbor in the clockwise sense is then rewired at
random, with probability $p$, to another vertex of the system,
chosen with some criterion. With probability $(1-p)$ the original
link is preserved. Self-connections and multiple connections are
prohibited. With this procedure, we have a regular lattice at
$p=0$, and progressively random graphs for $p>0$. The long range
links that appear at any $p>0$ trigger the small world
phenomenon. At $p=1$ all the links have been rewired, and the
result is similar to (though not exactly) a completely random
network. This algorithm should be used with caution, since it can
produce disconnected graphs. We have used only connected ones for
our analysis. Since links are neither destroyed nor created, the
resulting network has an average coordination number $2K$, equal
to the initial one.

We propose to chose the endpoint of the reconnected links
according to a probability distribution $\Theta (q)$ depending on
the topological distance $q$ between the two involved nodes in
the regular network (before any reconnection takes place). The
case with uniform probability is thus equivalent to the
Strogatz-Watts model. Other distributions may favor the
connections with nearby or with far away nodes. The values of $q$
run from 1 to the maximum distance allowed, namely $N/2K$. When
the node $i$ is to be rewired (according to the probability $p$)
we choose at random one among all the nodes at a certain distance
$q$ (taken with probability $\Theta (q)$). For example, as we
show below, by favoring reconnection with nearby nodes, we can
extend the interval of $p$ values where the small world behavior
is found, preserving the regular lattice high clusterization
$C(p,\Theta )$ as $L(p,\Theta )$ decreases. But besides the
macroscopical behavior of SW networks so generated, it is of
interest to explore their intrinsic structure. The results are
discussed in the following section.

\section{Numerical results and discussion}

\begin{figure}[h]
\centering
\resizebox{\columnwidth}{!}{\includegraphics{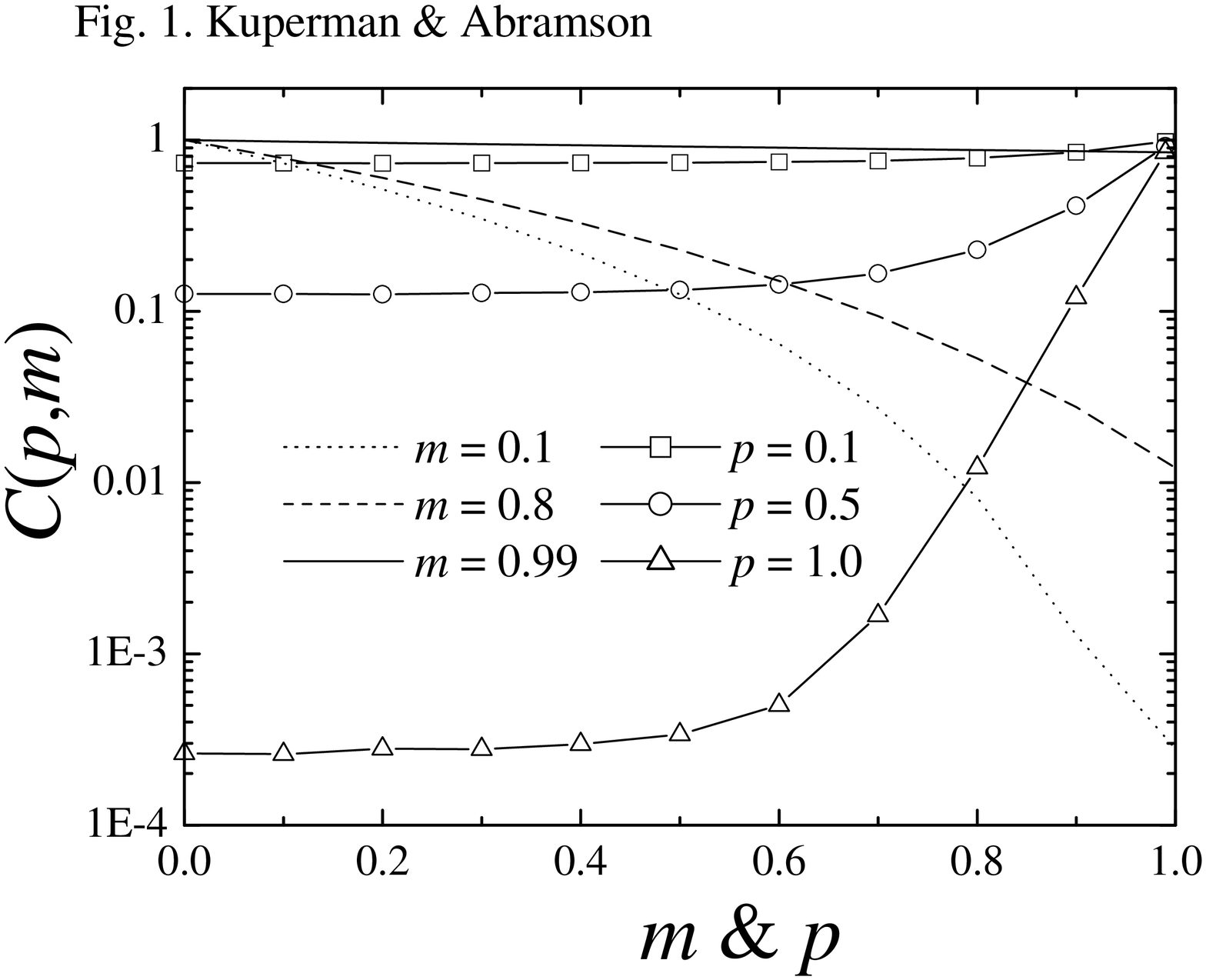}}
\caption{Average clusterization $C$ is displayed as a function of the
rewiring probability $p$, and of the exponent $m$, in the same
plot. Data are shown for three values of $p$ (lines and symbols),
and for three values of $m$ (lines), as indicated in the legend.
$N=10^5, K=10$.}
\label{cvsm+p}
\end{figure}

As an example of a generalization in the spirit of the previous
paragraph, we introduce only one family of SW networks taking
$\Theta \varpropto q^{-m}$, with $0<m<1$. The case $m=0$ gives a
uniform $\Theta (q)$ and is equivalent to already known cases.
For $m\ge 1$ a non integrable divergence at $q=0$ limits the
possibility of having a normalized probability distribution. Lets
start with the study of the average clusterization $C(p,m)$. In
Fig. \ref{cvsm+p}, we show two groups of curves. $C(p,m)$ as a
function of the rewiring parameter $p$ for three $m$ values is
shown with lines. $C(p,m)$ as a function of $m$ for three values
of $p$ is shown with lines and symbols. The system under study has
$N=10^{5}$ elements and $K=10$. We can see that as $m$ increases
the region of high clusterization moves toward higher values of
$p$. This means that the region in the $0\leq p\leq 1$ interval
where SW networks can be found is enlarged. The are other
relevant aspect derived from this fact that will be analyzed
later. This effect is much more evident in the second group of
curves, $C(p,m)$ \emph{vs} $m$. Moving along one of the curves,
which means fixing a value of $p$, for example $p=0.5$ (circles)
we can go from low clusterization at $m\approx0$ (uniform
distribution $\Theta$) to high clusterization as $m$ grows. All
the curves converge to the maximum clusterization when $m$
approaches 1.

\begin{figure}
\centering \resizebox{\columnwidth}{!}{\includegraphics{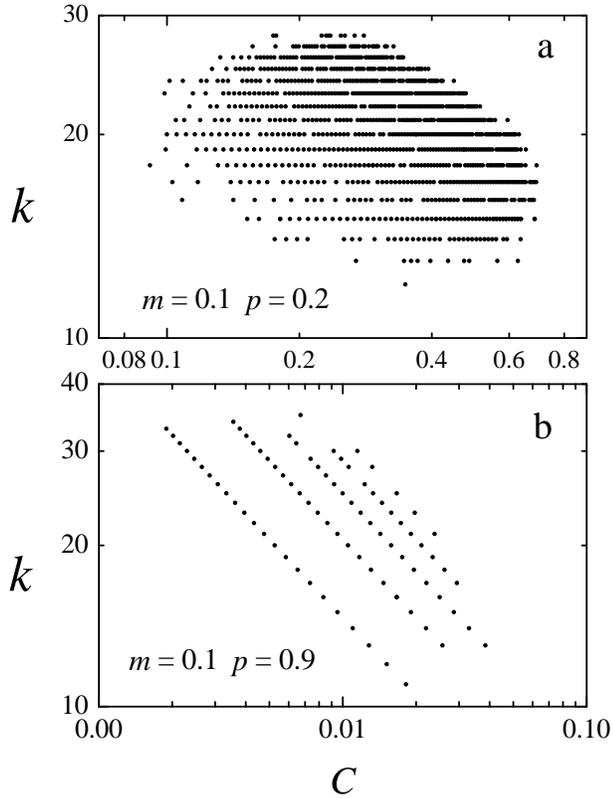}}
\caption{Connectivity \emph{vs} clusterization for each element in the
system. (a) $m=0.1$, $p=0.2$. (b) $m=0.1$, $p=0.9$. In both plots,
$N=10^5, K=10$. Except points with $c=0$ (not shown for reasons
of scale), all points are plotted.}
\label{connec}
\end{figure}

The other aspect of interest to be discussed is related to the
internal structure of the resulting networks. In recent papers
\cite{ama,albe}, it has been reported that different kinds of
behavior can be found in the accumulated connectivity
distribution of evolving networks. In the present model, as in
Strogatz and Watts', the connectivity presents a well defined mean
value, and an exponential decay at high values. This difference
is a consequence of the construction method. At the same time,
there is an interesting correlation between local clusterization
and connectivity values that is worth analyzing. In Fig.
\ref{connec} we show a plot of the connectivity vs. the
clusterization of each element in the system. It can be observed
that for each value of the connectivity there are elements with
widely different clusterization. This shows that a
characterization of this kind of networks in terms of the
connectivity alone is incomplete. It is remarkable that, at low
values of $p$, for each value of the connectivity, all (or almost
all) the allowed values of $C$ are present in the system. In
contrast, at high values of $p$, an emerging structure can be
observed. The elements fall into classes of clusterization. Each
class is characterized by a scale free relationship between
connectivity and clusterization. This is apparent in Fig.
\ref{connec}.b), where each class is represented by the dots
falling on straight lines in the log-log plot. In both plots, the
occupied region is upper bounded by an envelope with power law
decay. That  is, highly connected elements are restricted to a
range of low clusterizations, while lowly connected elements are
responsible for the preservation of the high clusterization of
the network. In qualitative terms we see that the network is
partitioned into small subnetworks composed mainly by elements of
low connectivity, while the long distance connection between each
subunit is accomplished by highly connected (popular but
otherwise lowly clustered) elements.

\begin{figure}[h]
\centering
\resizebox{\columnwidth}{!}{\includegraphics{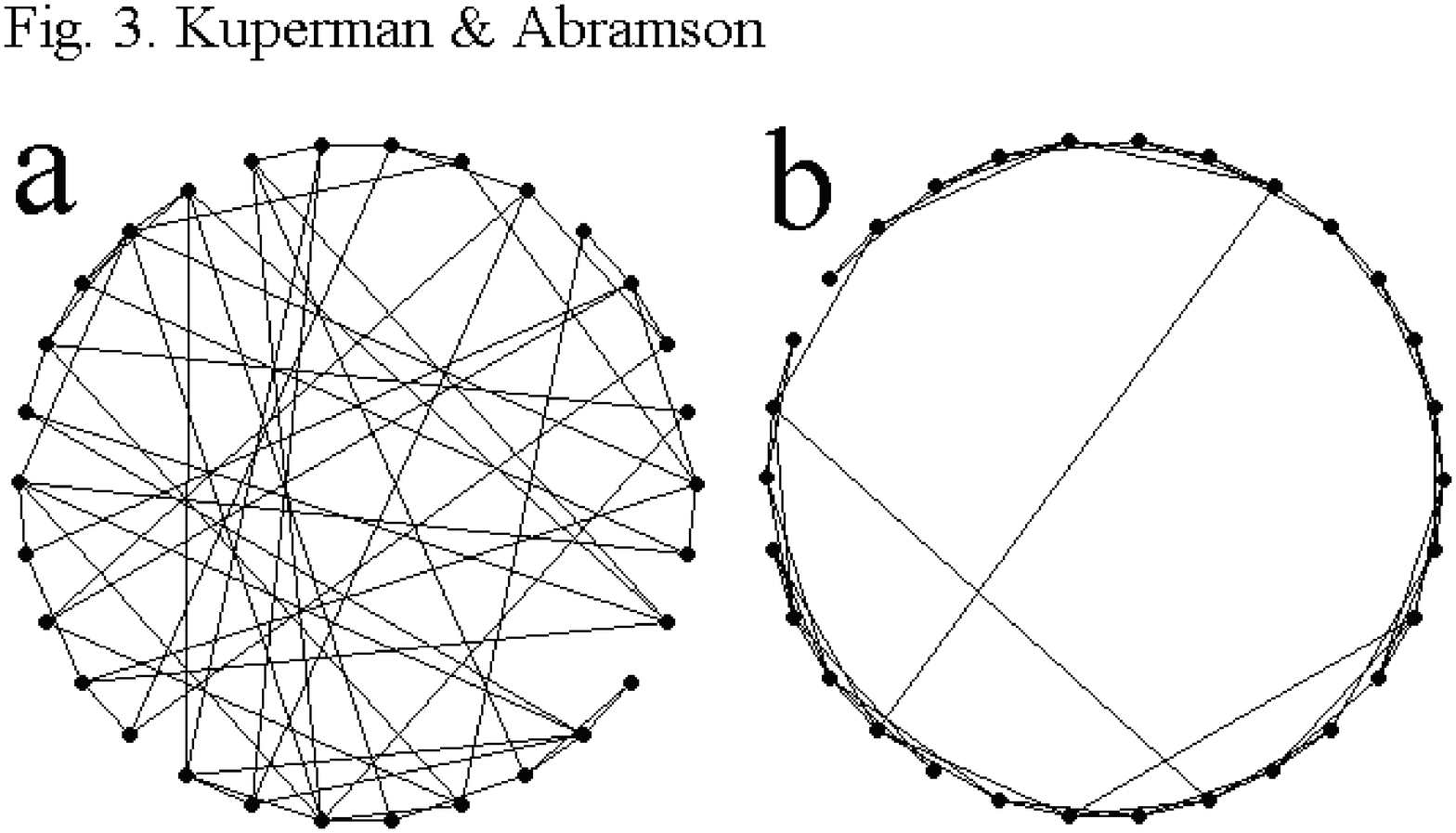}}
\caption{Two small worlds with 30 elements and $K=2$. They have the
same degree of rewiring but differing distribution $\Theta$. (a)
$p=0.5$, $m=0$. $C=0.11$ (b) $p=0.5$, $m=0.9$. $C=0.41$.}
\label{redes}
\end{figure}

When analyzing internally the structures of the obtained networks
we can see that several kinds of small world structures can
arise. An example for small systems ($N=30, K=2$) and two kind of
structures is shown in Fig. \ref{redes}. Both correspond to the
same value of the rewiring parameter $p=0.5$, meaning that
approximately one half of the links have been reconnected. In
Fig. \ref{redes}.b) we can see an organization of the links in two
levels: part of the rewired links have formed local subnetworks
of high clusterization, while some others are connecting these
structures between them. This is due to the fact that the
distribution $\Theta=q^{-0.9}$ favors links to near vertices. In
Fig. \ref{redes}.a), instead, $\Theta$ is uniform, there is no
preference in the reconnection process,  and no local structures
are formed. Case (b) could, at first sight, be confused with an
$m=0$ network with a value of $p$ significantly lower than the
current 0.5. However, in that case, much less links would have
been rewired. The average clusterization would be around 0.5
instead of 0.4. In a system with significantly greater $N$ and
$K$, one would expect a hierarchical organization in the
connectivity of subnetworks. This organization can be controlled
by the election of the distribution $\Theta$.

\begin{figure}
\centering \resizebox{\columnwidth}{!}{\includegraphics{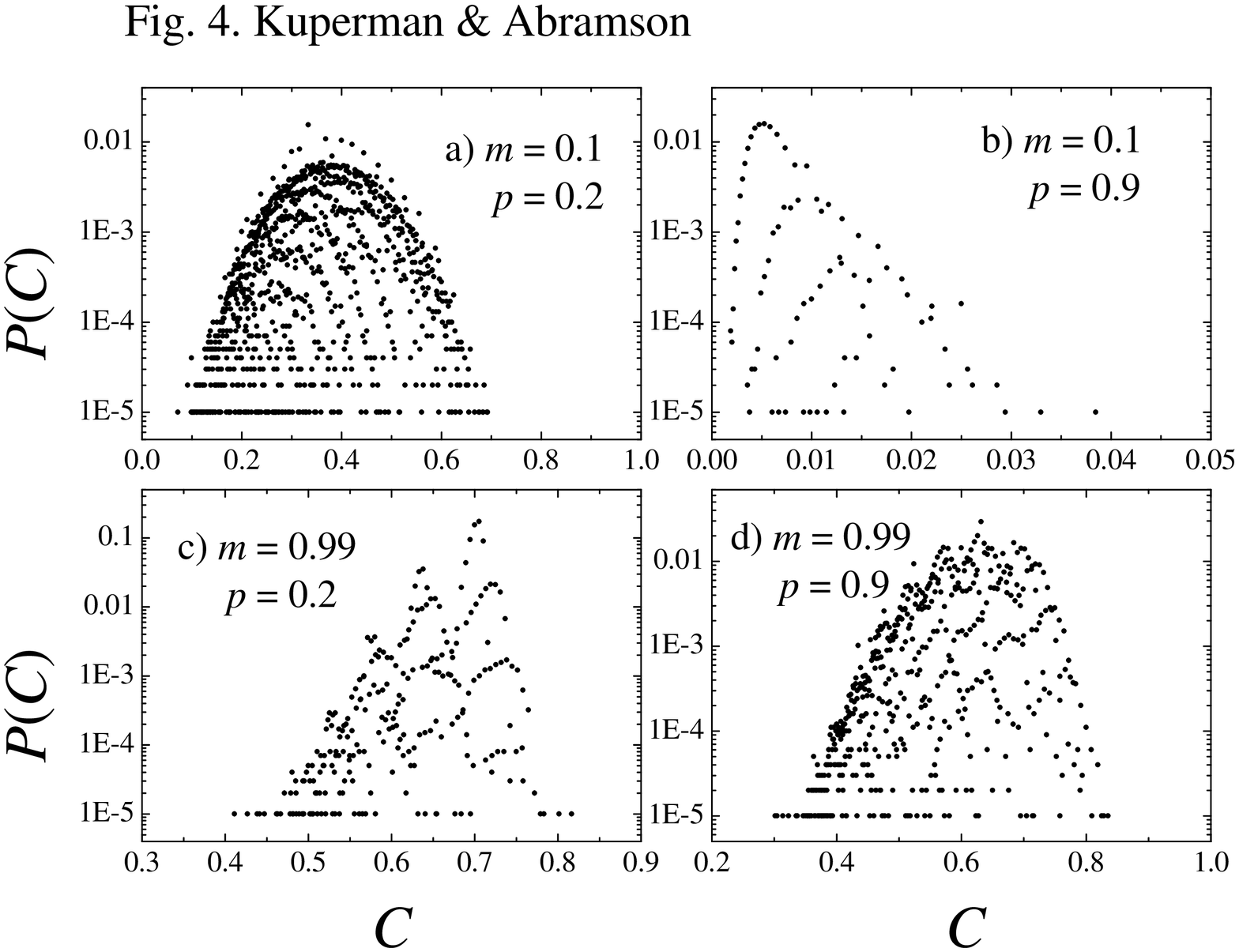}}
\caption{Distribution of the local clusterization. Values of $m$
and $p$ as indicated in the legends. In all plots, $N=10^5, K=10$.
Except points with $c=0$ (not shown for reasons of scale), all
points are plotted.}
\label{pvsc}
\end{figure}

Let us now turn to the distribution of the clusterization in the
system. In Fig. \ref{pvsc} we show histograms of the values of the
individual clusterization in the system, for four selected values
of $m$ and $p$. Each value of the local clusterization present in
the system (a discrete magnitude) is represented by a dot, at a
height proportional to its frequency. We can observe the incidence
of the parameters into the internal organization of the
clusterization of the network. The effect of the rewiring $p$ is
different at different values of $m$. It is interesting to observe
that each distribution is composed by a superposition of many
branches. Each one of these comprises a number of elements of
nearby clusterization.

\begin{figure}[t!]
\centering
\resizebox{\columnwidth}{!}{\includegraphics{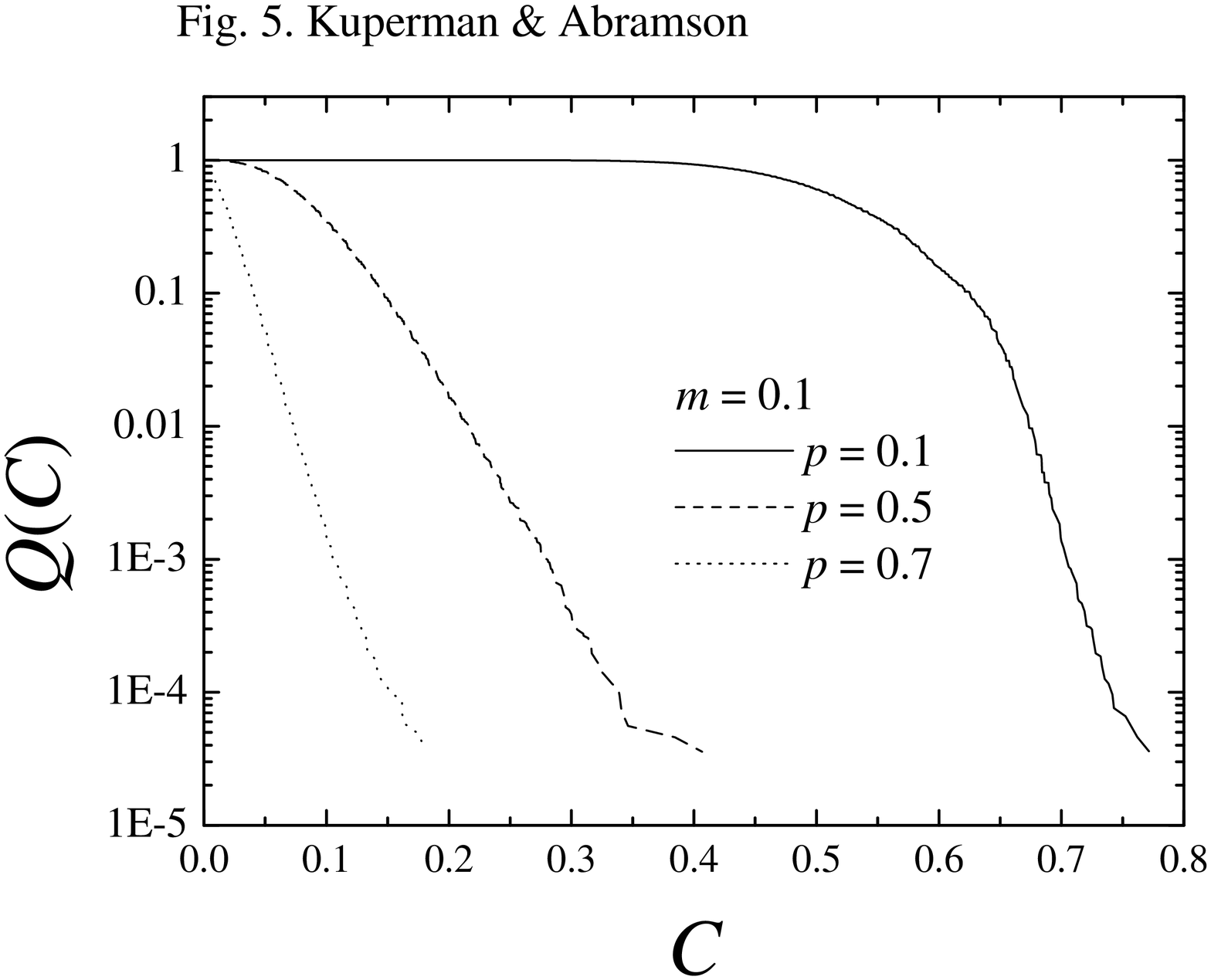}}
\caption{Accumulated probability of finding an element with clusterization $C$
or more. Data are shown for $m=0.1$, and three values of $p$, as
indicated in the legend. $N=10^5$ and $K=10$.}
\label{pac01}
\end{figure}

\begin{figure}[b!]
\centering
\resizebox{\columnwidth}{!}{\includegraphics{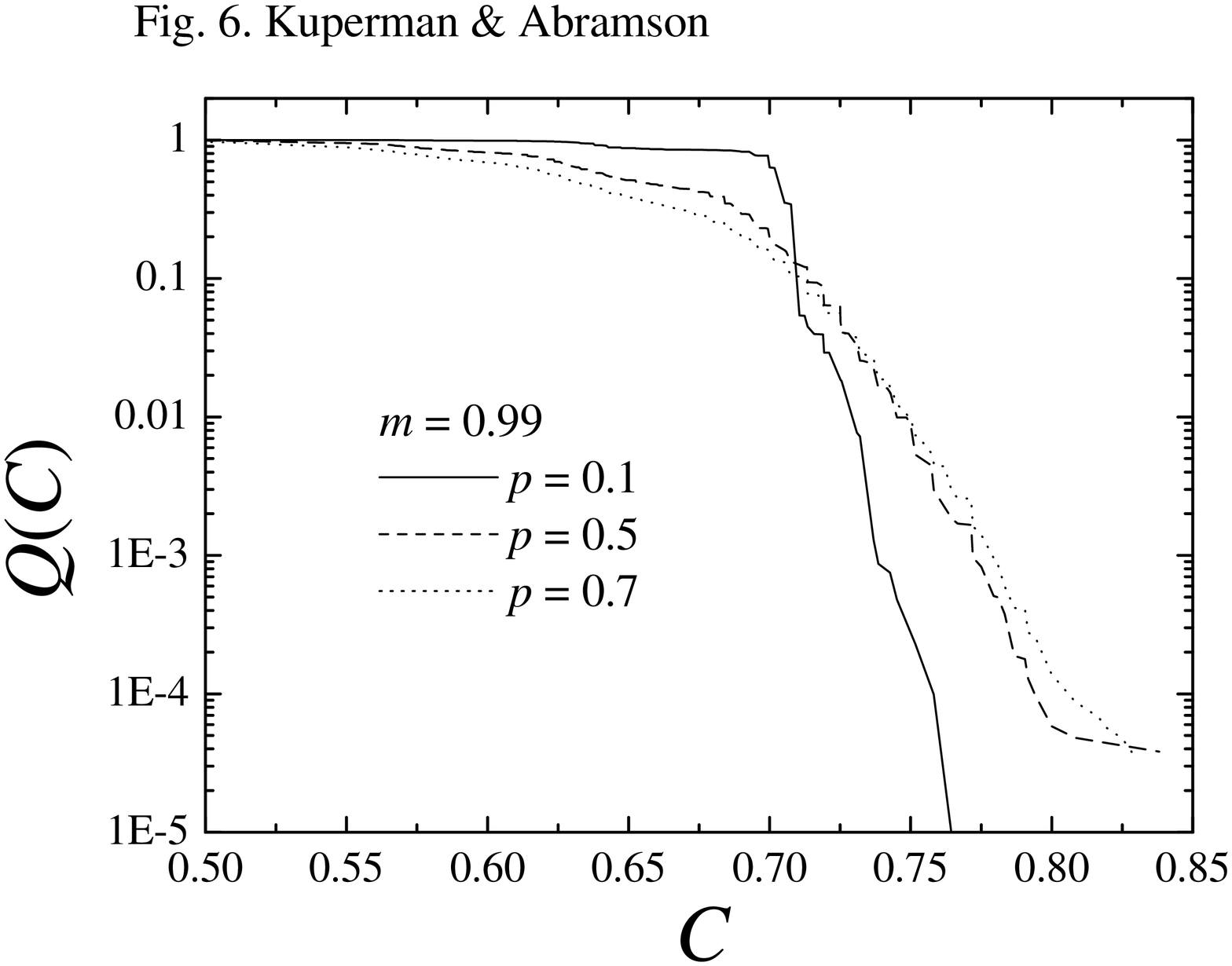}}
\caption{Accumulated probability of finding an element with clusterization $C$
or more. Data are shown for $m=0.99$, and three values of $p$, as
indicated in the legend. $N=10^5$ and $K=10$.}
\label{pac099}
\end{figure}

We calculate now the cumulative probability distribution of
clusterization $Q(C)=\sum_C^{1} P(C')$, that represents the
probability that an element has clusterization $C$ or greater. In
figures \ref{pac01} and \ref{pac099} we show the accumulated
clusterization $Q(C)$ for three values of the rewiring parameter
$p$. Fig. \ref{pac01} corresponds to a slightly non uniform
distribution $\Theta=q^{-0.1}$ while Fig. \ref{pac099} corresponds
to a highly non uniform distribution $\Theta=q^{-0.99}$. First,
observe (in any of the figures) that different values of $p$
produce different behaviors in the accumulated distribution. In
Fig. \ref{pac01} we have $m=0.1$. This slight departure from the
uniformity (Strogatz and Watts' model) produces a slower decay of
$Q$ for growing $p$, appreciable as a change of the curvature in
the log plots. The fact that $\Theta=q^{-0.99}$ generates networks
with many highly clustered elements (independently of $p$) can be
seen in Fig. \ref{pac099} as a plateau that reaches high values
of $C$. Even if the three distributions in Fig. \ref{pac099}
decay in similar way, they arise from distributions $P(C)$ as
differing as those shown in Fig. \ref{pvsc} (bottom row). Note
also in Figs. \ref{pac01} and \ref{pac099} that the effect of a
$\Theta$ that favors high clusterization can be seen even at low
levels of disorder as $p=0.1$. This can be appreciated by
comparing the curves corresponding to $p=0.1$ in Figs.
\ref{pac01} and \ref{pac099} (full lines), where the plateau in
$Q(C)$ shows that almost all the values of clusterization are
above $0.4$ in the first case, and above $0.7$ in the second.

We have used $\Theta=q^{-m}$ as an example of a distribution used
to generate the rewiring. The present scheme is rather flexible to
produce a variety of clusterization distributions and topologies,
through the election of a proper distribution $\Theta$. This fact
may be relevant in the modelling of real networks. At the present
time several social phenomena have been modelled on the basis of a
SW network \cite{abra,kup,pastor}.


\end{document}